# Ultrafast dynamics and sub-wavelength periodic structure formation following irradiation of GaAs with femtosecond laser pulses

A. Margiolakis,[1,2] * G. D. Tsibidis,[3] K. M. Dani,[2] and G. P. Tsironis[1,3]

[1]*Department of Physics, University of Crete, P. O. Box 2208, 71003 Heraklion, Greece*

[2]*Femtosecond Spectroscopy Unit, Okinawa Institute of Science and Technology Graduate University, 1919-1 Tancha, Onna-son, Kunigami, Okinawa904-495, Japan*

[3] *Institute of Electronic Structure and Laser (IESL), Foundation for Research and Technology (FORTH), N. Plastira 100, Vassilika Vouton, 70013, Heraklion, Crete, Greece*

A theoretical investigation of the ultrafast processes and dynamics of the excited carriers upon irradiation of GaAs with femtosecond (fs) pulsed lasers is performed in conditions that induce material damage and eventually surface modification of the heated solid. A parametric study is followed to correlate the produced transient carrier density with the damage threshold for various pulse duration values $\tau_p$ (it increases as $\sim \tau_p^{0.053 \pm 0.011}$ at relatively small values of $\tau_p$ while it drops for pulse durations of the order of some picoseconds) based on the investigation of the fundamental multiscale physical processes following fs-laser irradiation. Moreover, fluence values for which the originally semiconducting material demonstrates a metallic behaviour are estimated. It is shown that a sufficient number of carriers in the conduction band are produced to excite Surface Plasmon (SP) waves that upon coupling with the incident beam and a fluid-based surface modification mechanism lead to the formation of sub-wavelength periodic structures orientated perpendicularly to the laser beam polarization. Experimental results for the damage threshold and the frequencies of induced periodic structures show a good agreement with the theoretical predictions.

## I. INTRODUCTION

Over the past decades, laser-based material processing with ultra-short pulsed laser sources has received considerable attention due to its important technological applications, in particular in industry and medicine [1-9]. Rapid energy delivery and reduction of the heat-affected areas are the most pronounced advantages of the technique compared to effects induced by longer pulses [10], which reflect the merit of the method as a potential tool for laser-assisted fabrication at micro- and nano-scales.

One type of surface modification, the so-called laser-induced periodic surface structures (LIPSS) on solids have been studied extensively for linearly polarized beams. In recent works, it was also shown that it is possible to produce even more functional surfaces by using more complex polarisation states. It turns out that the morphological features of structures provide unique opto-mechanical properties that can be used in many applications [11-14].

Previous theoretical approaches or experimental observations related to the understanding of the underlying physical mechanism for the formation of these structures were performed in a variety of conditions [15-27]. The most representative types of LIPSS that have been explored are the usually termed low-spatial-frequency (LSFL) ripples, high-spatial-frequency (HSFL) ripples, grooves and spikes. In case of metallic and semiconducting materials, ripples are formed at low, microgrooves at intermediate [28] and quasi-periodic arrays of microspikes [29,30] at high number of pulses (*NP*) or fluence. LSFL have spatial periods of the order of the laser wavelength $\lambda_L$. In most materials, they are oriented perpendicularly to the laser beam polarization. In dielectrics, LSFL were observed either perpendicular or (for very large band gap materials) parallel to the beam polarization. By contrast, grooves are supra-wavelength structures and are orientated parallel to the polarization of the laser beam [21,23,31] .

To explain the underlying physical origin of LIPSS formation, it is important to note that following irradiation with ultrashort pulses, a series of multiphysical phenomena take place [32-36]. With respect to the formation of ripples, various mechanisms have been proposed to account for the production of periodic structures: interference of the incident wave with an induced scattered wave [16,18,21], or with a surface plasmon wave (SP) [17,20,37-40], or due to self-organisation mechanisms [41].

While the precise physical mechanism for the origin of LIPSS is still debatable, one process that undoubtedly occurs is a phase transition that eventually leads to a surface modification. Physical mechanisms that lead to surface modification have been explored both theoretically and experimentally [9,17,20,21,31,36,40,42-47] and it is evident that a precise determination of the morphological changes upon irradiation requires a thorough investigation of phase transitions and resolidification process.

On the other hand, femtosecond laser interaction involves several complex phenomena, including energy absorption, photon-ionisation processes, electron excitation, electron-relaxation processes, phase transitions and/or thermomechanical effects, resolidification and mass ejection. In principle, the laser beam parameters (wavelength, pulse duration, fluence, number of pulses, angle of incidence and beam polarisation state) determine the onset of the surface modification as energy absorption, electrodynamical effects and relaxation processes are critical to the material heating. The complexity of the processes and ionisation mechanisms is material dependent as the laser source is used to excite, firstly, electrons/carriers from the valence to the conduction band before the energy is transferred into the lattice system [48]. One characteristic, though, that influences the thermal response of the material is the amount of the absorbed energy which is also closely related the electron excitation level (i.e. reflectivity) and dynamics.

To fully understand the ultrafast dynamics of the excited carriers upon irradiation with ultrashort pulses, it is important to perform a thorough analysis of the influence of the laser parameters on the thermal response of the material. Theoretical models that describe the fundamentals of laser-matter interaction for various types of materials (i.e. metals, semiconductors, dielectrics) and experimental studies successfully provide a detailed analysis of the physical mechanisms behind a plethora of structural effects (i.e. production of craters, evaluation of damage thresholds, LIPSS formation) [20,48-52].

Nevertheless, although for many semiconducting materials, the physical mechanism that describes ultrafast dynamics is well-established and the theoretical model works efficiently in various conditions (that lead to high excited carrier densities, $\sim 10^{20}$ - $10^{22}$ cm$^{-3}$), there is still a missing picture for some types of semiconductors such as GaAs [33,53-56]. More specifically, GaAs is characterised by a higher electron mobility and higher thermal stability than Si, and it has a direct band gap which makes it more efficient in absorbing and emitting light. Thus, this material has a better performance in solar-harvesting energy-related applications [57] or terahertz antennas [58]. A complete understanding of the ultrafast electron and lattice dynamics for large number of excited carriers ($> 10^{20}$ cm$^{-3}$), close and beyond the damage threshold (i.e. associated to the fluence that induces melting) of GaAs, will allow the identification of the dominant processes in this regime. Furthermore, behaviour of the material in conditions that lead to highly excited carrier densities is expected to allow an optimisation of laser-based micromachining of GaAs and produce morphologies (such as LIPSS) with impressive properties for the aforementioned applications. The elucidation of these issues is of paramount importance not only to understand further the underlying physical mechanisms of laser-matter interactions and ultrafast electron dynamics but also to associate the resulting thermal effects with the surface response. Therefore, there is a growing interest to reveal the physics of the underlying processes from both a fundamental and application point of view.

To this end, we present an extension of the well-established theoretical model that describes ultrafast dynamics in semiconductors [20,32,36,59-61], to account firstly, for excitation and electron-phonon relaxation upon heating GaAs with ultrashort pulsed lasers (Section II). To the best of our knowledge, a theoretical investigation of the fundamental multiscale processes has not been performed for GaAs. The theoretical framework is coupled to a module that accounts for the formation of SP-generated LSFL ripples by predicting the laser conditions for the production of sufficiently high density of excited carriers. As the laser conditions of the simulations lead to a phase transition, the role of fluid dynamics in the modulation of the surface profile is briefly explored. Section III explains the details of the numerical algorithm used and the adaptation of the model to GaAs. The details of an experimental protocol that has been developed are given in Section IV while a systematic analysis of results and validation of the theoretical model are presented in Section V by estimating the damage threshold and ripple periodicities. Concluding remarks follow in Section VI.

## II. THEORETICAL MODEL

### a. Energy and Particle Balance equations

During laser irradiation of a semiconducting material, various physical processes occur on a femtosecond timescale. As excitation of GaAs is performed through a laser beam of $\lambda_L = 800$ nm corresponding to photon energy equal to 1.55eV that is higher than the band gap of the material ($\sim 1.42$ eV, at 300 K), it is assumed that one- and two-photon absorption mechanisms are sufficient to excite carriers from the valence to the conduction band while higher order photon processes are less likely to occur. On the other hand, (linear) free carrier photon absorption can increase the electron energy (but not the number of the excited carriers) while Auger recombination and impact ionization processes lead to decrease or increase of the carriers in the conduction band, respectively.

To describe the carrier excitation and relaxation processes, the relaxation time approximation to Boltzmann's transport equation [20,32,36,59-61] is employed to determine the spatial ($\vec{r} = (x,y,z)$) and temporal dependence ($t$) of the carrier density number, carrier energy and lattice energy; based on this picture the following set of coupled (nonlinear) energy and particle balance equations are used to derived the evolution of the carrier density number $N_e$, carrier temperature $T_c$ and lattice temperature $T_L$

$$C_c \frac{\partial T_c}{\partial t} = -\frac{C_c}{\tau_e}(T_c - T_L) + S(\vec{r},t)$$

$$C_L \frac{\partial T_L}{\partial t} = \vec{\nabla}\bullet(K_L \vec{\nabla} T_L) + \frac{C_c}{\tau_e}(T_c - T_L) \qquad (1)$$

$$\frac{\partial N_e}{\partial t} = \frac{\alpha_{SPA}}{\hbar \omega_L} I(\vec{r},t) + \frac{\beta_{TPA}}{2\hbar \omega_L} I^2(\vec{r},t) - \gamma N_e^3 + \theta N_e - \vec{\nabla}\bullet\vec{J}$$

where $C_c$ ($C_L$) is the carrier (lattice) heat capacity, $k_e$ ($k_h$) is the heat conductivities of the electron (holes), $\hbar\omega_L$ stands for the photon energy (~1.55eV for $\lambda_L$ = 800 nm), $\alpha_{SPA}$ and $\beta_{TPA}$ correspond to the single and two-photon absorption coefficients, respectively, $\gamma$ is the coefficient for Auger recombination, $\theta$ is the impact ionization coefficient, and $\tau_e$ is the carrier-phonon energy relaxation time. It is noted, that the expression for Si was used to approximately estimate $\tau_e$ as there is not any reported relevant value. Through this expression, the significance of carrier density dependence of the relaxation time is recognised. Other quantities that need to be evaluated are the carrier current density $\vec{j}$ and the heat current density $\vec{w}$

$$S(\vec{r},t) = (\alpha_{SPA} + \alpha_{FCA})I(\vec{r},t) + \beta_{TPA} I^2(\vec{r},t) - \vec{\nabla}\bullet\vec{W}$$
$$- \frac{\partial N_e}{\partial t}(E_g + 3k_B T_e) - N_e \left( \frac{\partial E_g}{\partial T_L} \frac{\partial T_L}{\partial t} + \frac{\partial E_g}{\partial N_e} \frac{\partial N_e}{\partial t} \right)$$
$$\vec{W} = (E_g + 4k_B T_e)\vec{J} - (k_e + k_h)\vec{\nabla}T_e$$
$$\vec{J} = -D\left( \vec{\nabla}N_e + \frac{N_e}{2k_B T_e}\vec{\nabla}E_g + \frac{N_e}{2T_e}\vec{\nabla}T_e \right) \qquad (2)$$
$$D = \frac{2k_B T_e}{e} \frac{\mu_e^0 \mu_h^0}{\mu_e^0 + \mu_h^0}$$
$$C_c = 3N_e k_B + N_e \frac{\partial E_g}{\partial T_e}$$

where $D$ stands for the ambipolar carrier diffusivity, $\mu_e^0$ and $\mu_h^0$ are the electron/hole mobilities (8500 cm$^2$/Vs and 400 cm$^2$/Vs, respectively, for GaAs [62]) and $e$ is the electron charge (all values of all parameters and coefficients used in this work are presented in Table I.). It needs to be emphasized that it was assumed that the carrier system is non-degenerate (it follows a Maxwell-Boltzmann distribution) [59]. Previous reports (for Si) showed that estimation of damage thresholds or carrier density values do not differ significantly if this simplification is ignored [60]. Similarly, the influence of the carrier heat diffusion is ignored (we set, for simplicity, $k_e \simeq k_h \simeq 0$) as carrier diffusion has in general, little impact on the creation of electron-hole carriers [59,60]. This approximation is valid for Silicon and it can be assumed that it holds true for GaAs ([59,60], given the computed values of $k_e$, $k_h$ are also smaller than those of Si [63]).

The laser intensity $I(\vec{r},t)$ in Eqs.(1-2) is obtained by considering the propagation loss due to one-, two- photon and free carrier absorption, respectively [64]

$$\frac{\partial I(\vec{r},t)}{\partial z} = -(\alpha_{SPA} + \alpha_{FCA})I(\vec{r},t) - \beta_{TPA} I^2(\vec{r},t) \qquad (3)$$

assuming that the laser beam is Gaussian, both temporally and spatially, and the transmitted laser intensity at the incident (flat) surface is expressed in the following form

$$I(x,y,z=0,t) = \frac{2\sqrt{\ln 2}E_p(1-R(z=0,t))}{\sqrt{\pi}\tau_p} e^{-\left(\frac{2(x^2+y^2)}{R_0^2}\right)} e^{-4\ln 2\left(\frac{t-t_0}{\tau_p}\right)^2} \qquad (4)$$

where $E_p$ is the fluence of the laser beam and $\tau_p$ is the pulse duration (i.e. full width at half maximum), $R_0$ is the irradiation spot-radius (distance from the centre at which the intensity drops to $1/e^2$ of the maximum intensity, and $R$ is the reflectivity while irradiation under normal incidence was assumed.

The computation of the free carrier absorption coefficient and the reflectivity are derived from the dielectric constant of the material (assuming also corrections due to band and state filling [65]), $\varepsilon'$,

$$\varepsilon' = 1 + (\varepsilon_{un} - 1)\left(1 - \frac{N_e}{N_v}\right) - \frac{e^2 N_e}{\varepsilon_0 \omega_L^2}$$
$$\times \left[ \frac{1}{m_{e-cond}^*\left(1 + i\frac{1}{\omega_L \tau_{col}}\right)} + \frac{1}{m_{h-cond}^*\left(1 + i\frac{1}{\omega_L \tau_{col}}\right)} \right] \qquad (5)$$

where $\varepsilon_{un}$ is the dielectric constant of the unexcited material at $\lambda_L$ = 800 nm ($\varepsilon_{un}$ = 13.561+ 0.63105i) [66], $m_{e-cond}^*$ = 0.067 $m_{e0}$ $m_{h-cond}^*$ = 0.34 $m_{e0}$ are the optical effective masses of the carriers [67] for conductivity calculations, $m_{e0}$ is the electron mass, $\varepsilon_0$ is the permittivity of vacuum, $N_v$ corresponds to the valence band carrier density (~5×10$^{22}$ cm$^{-3}$) and $\tau_{col}$ stands for the carriers (electron-hole) collision time. It is noted that optical effective masses are taken to be constant for GaAs and the excitation conditions do not alter them. Similar assumptions were made for Silicon or Germanium that demonstrate that constant values yielded an adequate description of dynamics and morphological effects [20,32,36,59,60].

Despite estimates in other reports that assume a dynamical variation of $\tau_{col}$ [60], for the sake of simplicity, a constant value, $\tau_{col}$ ~ 10 fs is assumed in the simulations as for Si [36]. The reflectivity and free carrier absorption coefficients are given by the following expressions (i.e. for a flat surface)

$$\alpha_{FCA}(x,y,z,t) = \frac{2\omega_L k}{c}$$

$$R(x,y,z=0,t) = \frac{(1-n)^2 + k^2}{(1+n)^2 + k^2} \quad (6)$$

where *c* stands for the speed of light while *n* and *k* are the refractive index and extinction coefficient of the material, respectively. By contrast, reflectivity for a nonflat profile depends on the corrugation of the morphology and it denoted by *R(x,y,surface,t)* (computation of *R(x,y,surface,t)* is described in Section III) while Eq.4 is replaced with

$$I(x,y,surface,t) = \frac{2\sqrt{\ln 2}E_p(1-R(x,y,surface,t))}{\sqrt{\pi}\tau_p} e^{-\left(\frac{2(x^2+y^2)}{R_0^2}\right)} e^{-4\ln 2\left(\frac{t-t_0}{\tau_p}\right)^2} \quad (7)$$

**b. Surface Plasmon Excitation**

As explained in the Introductory section, the excitation of SP and its interference with the incident beam constitutes one of the dominant mechanisms that aim to explain the formation of LSFL [20,35,36,68,69]. Nevertheless, coupling of the incident beam with SP modes is possible only if there is an initial corrugation (for example, surface defects [20]) or via index matching [68] and grating coupling [70]. Due to an inhomogeneous deposition of the laser energy on the semiconductor as a result of the exposure to a linearly polarised Gaussian-shape beam, the surface of material is not expected to be perfectly smooth after resolidification following irradiation with *NP*=1 (crater [20]); hence, further irradiation of the non-planar profile will give rise to a SP which is capable to couple with the incident beam yielding a resultant periodically modulated intensity ($\sim \exp(i\vec{k}_s \cdot \vec{r})$) [20] ($I_{total}(\vec{r},t) \sim \langle |\vec{E}_{incident} + \vec{E}_{SP}|^2 \rangle$ [20,71]. Thus, form this expression, it is evident that the interference of the incident and the surface waves (SP) leads to a periodically modulated intensity profile.

According to theoretical predictions and experimental studies, the interference of the incident and the surface wave results in the development of periodic ripples (LSFL) with orientation *perpendicular* to the electric field of the laser beam [15-27]; this is generated, firstly, by the development of the aforementioned spatially periodic energy distribution that yields a periodic distribution of the electron temperature field. Upon relaxation due to electron-phonon scattering, the characteristic spatial modulation will be projected on the lattice system and fluid dynamics (when the material undergoes a phase transition) leading eventually to a rippled profile when the resolidification process ends [20,24,25,72].

The dispersion relation for the excitation of SP is derived by the boundary conditions (continuity of the electric and magnetic fields at the interface between a metallic and dielectric material) ($\varepsilon_d = 1$) for a flat surface (*NP*=1). Therefore, a requirement for a semiconductor to obey the above relation and conditions [20,68] is that *Re(ε')<-1* and the computed SP wavelength $\lambda_S$ is given by

$$\lambda_S = \frac{\lambda_L}{\operatorname{Re}\sqrt{\left(\frac{\varepsilon'\varepsilon_d}{\varepsilon'+\varepsilon_d}\right)}} \quad (8)$$

The condition *Re(ε')<-1* and Eqs.(5,8) can be used to derive the range of values of the excited carrier densities that lead to SP excitation. It is evident that carrier densities larger than ~1.5 × 10²¹ cm⁻³ lead to excitation of SP (Fig.1). Although, Eq.(8) can be used to calculate the SP and, eventually, the ripple periodicity, there exists a discrepancy between the experimental observations and theoretical prediction. More specifically, special attention is required when the surface profile changes from a flat to a periodically corrugated one. Previous studies reported that a systematic analysis of the distribution of the electric field of the laser beam shows that there exists a correlation of the SP wavelength with the depth of the periodic structure that leads to the best SP-laser coupling and therefore Eq.8 is no longer valid [17,24,73] (see discussion in Sections III, V).

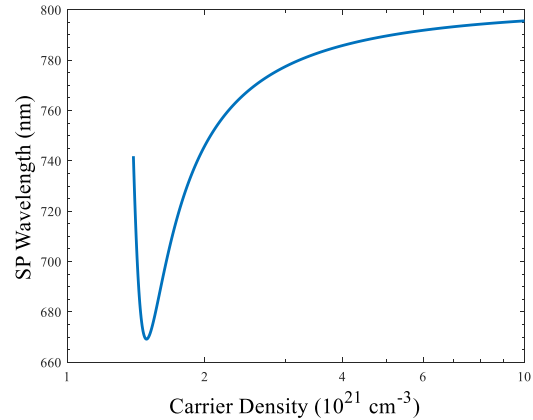

FIG. 1 (Color Online): Surface plasmon wavelength as a function of the excited carrier densities when the SP excitation condition is satisfied.

As pointed out above, although Eq.8 indicates the SP excitation condition for *NP*=1, a coupling of the SP with the incident beam requires a corrugation/defect on the surface; given that for *NP*=1, a corrugation is not present, an interference between a SP and the incident beam is not possible and therefore ripples cannot be produced. By contrast, the modified profile (following surface modification after *NP*=1) produces the initial corrugation required to both excite SP and induce coupling with the incident beam [68]. This will lead to the required inhomogeneous energy deposition that eventually leads to a periodic structure formation [20,37,74] through a periodic modualation of the intensity (Sections III,V).

### c. Fluid Dynamics and material removal

Due to the need for the consideration of a phase transition for the description of an induced morphological change, fluid dynamics is introduced. The material that undergoes melting is assumed to be an incompressible Newtonian fluid and its dynamics is described by the following equations [20]:

(i). for the mass conservation (incompressible fluid):

$$\vec{\nabla}\cdot\vec{u} = 0 \tag{9}$$

(ii). for the energy conservation:

$$C_L^{(m)}\left(\frac{\partial T_L^{(m)}}{\partial t} + \vec{\nabla}\cdot\left(\vec{u}T_L^{(m)}\right)\right) = \vec{\nabla}\cdot(K_L^{(m)}\vec{\nabla}T_L^{(m)}) + \frac{C_c}{\tau_e}(T_c - T_L) \tag{10}$$

where $K_L^{(m)}$ is the thermal conductivity of the molten material. The presence of a liquid phase requires a modification of the second of Eq.1 to incorporate convection. This means that while the second equation in Eq.1 that describes the lattice temperature evolution still holds for material which is in the solid phase, an appropriate modification is required to account for the transport/convection of the fluid [20]. Furthermore, an additional term with a *δ*-function is presented in Eq.10 to describe a smooth transition from the liquid-to-solid phase

$$C_L^{(m)}\left[\frac{\partial T_L^{(m)}}{\partial t} + \vec{\nabla}\cdot\left(\vec{u}T_L^{(m)}\right)\right] + L_{melt}\delta\left(T_L^{(m)} - T_{melt}\right)\frac{\partial T_L^{(m)}}{\partial t} = \vec{\nabla}\cdot(K_L^{(m)}\vec{\nabla}T_L^{(m)}) + \frac{C_c}{\tau_e}(T_c - T_L) \tag{11}$$

(iii). for the momentum conservation:

$$\rho_L^{(m)}\left(\frac{\partial \vec{u}}{\partial t} + \vec{u}\cdot\vec{\nabla}\vec{u}\right) = \vec{\nabla}\cdot\left(-P\mathbf{1} + \mu\left(\vec{\nabla}\vec{u}\right) + \mu\left(\vec{\nabla}\vec{u}\right)^T\right) \tag{12}$$

where $\vec{u}$ is the velocity of the fluid, $\mu$ is the liquid viscosity, *P* is the total pressure (hydrodynamical and recoil) and $C_L^{(m)}$ stands for the heat capacity of the liquid phase. Vapour ejected creates a back (recoil) pressure on the liquid free surface which in turn pushes the melt away in the radial direction. The recoil pressure and the surface temperature are usually related according to the equation [75,76]

$$P_r = 0.54 P_0 \exp\left(L_v \frac{T_L^S - T_b}{RT_L^S T_b}\right) \tag{13}$$

where $P_0$ is the atmospheric pressure ($\sim 10^5$ Pa), $L_v$ is the latent heat of evaporation (of the order of $10^7$ kJ/m$^3$ as for Silicon and Germanium [77]), $T_b$ stands for the boiling temperature, *R* is the universal gas constant, and $T_L^S$ corresponds to the surface temperature. The surface tension in pure molten GaAs decreases with growing melt temperature (Table I), which causes an additional depression of the surface of the liquid closer to the maximum value of the beam while it rises elsewhere. Hence, spatial surface tension variation induces stresses on the free surface and therefore a capillary fluid convection is produced. Thus, a surface tension related pressure is derived, $P_\sigma$ which is expressed as $P_\sigma \sim \sigma$.

(iv) shallow water equations:

Due to the phase transition, the dynamics of the molten material will determine the surface profile when solidification terminates. Thus, the generated ripple height is calculated from the Saint-Venant's shallow water equation [78]

$$\frac{\partial H(\vec{r},t)}{\partial t} + \vec{\nabla}\cdot\left(H(\vec{r},t)\vec{u}\right) = 0 \tag{14}$$

where $H(\vec{r},t)$ stands for the melt thickness which evolves and provides the final surface morphology upon resoldification.

The presence of an intermediate zone that contains material in both phases when the solid undergoes a phase transition will complicate the description of flow dynamics (presence of a mushy zone [79]). Nevertheless, to avoid complexity of the solution of the problem and given the small width of the two phase region with respect to the size of the affected zone a different approach will be pursued where a mushy zone is neglected and transition from to solid-to-liquid is indicated by a smoothened step function of the thermophysical quantities. In previous reports, the role of the recoil pressure and critical temperatures were also taken into account to model ablation (material removal) conditions [20] (i.e. boiling [25] or critical point of the material [20] yielding a lattice temperature of the order of $\sim (2.5\text{-}3) \times T_{melt}$ or equivalently $\sim 0.90\ T_{cr}$ (see [20,80-82] and [63]). However, in the current work, due to the lack of knowledge of relevant parameters and for the sake of simplicity, a lattice temperature of the material equal to approximately $T_{rem} = 3 \times T_{melt}$ is assumed to lead to the onset of mass removal. Furthermore, $T_b$ is taken to be equal to ~2.5 × $T_{melt}$ (approximately similar to the experimentally confirmed relation for Silicon or Germanium).

## III. NUMERICAL SOLUTION

Due to the inherent complexity and highly nonlinear character of the equations in Section II, an analytical solution is not feasible and therefore, a numerical approach is pursued. Numerical simulations have been performed using a finite difference method while the discretization of time and space has been chosen to satisfy the Neumann stability criterion. Furthermore, it is assumed that on the boundaries, von Neumann boundary conditions are satisfied and heat losses at the front and back surfaces of the material are negligible. The initial conditions are $T_e(t=0) = T_L(t=0) = 300$ K, and $N_e = 10^{12}$ cm$^{-3}$ at $t=0$. The parameters for GaAs used in the simulation are summarised in Table I.

The values of the laser beam used in the simulation are: The (peak) fluence is $E_p \left( \equiv \sqrt{\pi} \tau_p I_0 / (2\sqrt{ln2}) \right)$, where $I_0$ stands for the peak value of the intensity while $R_0$ in Eq.4 is taken to be equal to 20 μm. The wavelength of the beam is $\lambda_L$=800 nm. A common approach followed to solve similar problems is the employment of a staggered grid finite difference method which is found to be effective in suppressing numerical oscillations. Unlike the conventional finite difference method, temperatures ($T_c$ and $T_L$), carrier densities ($N_e$), pressure ($P$) are computed at the centre of each element while time derivatives of the displacements and first-order spatial derivative terms are evaluated at locations midway between consecutive grid points. Similarly, the carrier current density $\vec{j}$ ($x, y, z, t$) and the heat current density $\vec{w}$ ($x, y, z, t$) are evaluated at the above locations rather than at the centres of each element. On the other hand, the horizontal and vertical velocities are defined in the centres of the horizontal and vertical cells faces, respectively (for a more detailed analysis of the numerical simulation conditions and the methodology towards the description of fluid dynamics, see Ref.[20]). It is also assumed that, for $NP$=1, $\vec{j}$ ($z=0, t$) and $\vec{w}$ ($z, t$) at the front ($z=0$) or the other end of the irradiated region ($z=10$ μm) vanish. By contrast, for $NP$>1 (for which the surface profile is modified), some modification is required: For irradiation with $NP$>1, the incident beam is not always perpendicular to the modified profile, therefore the surface geometry influences the spatial distribution of the deposited laser energy. For example, the laser irradiation reflected from the profile slopes can lead to light entrapment between the formed structures. Similarly, energy absorption is altered if incident beam irradiates at zero or different angle and therefore energy absorption varies on the slopes. Typical Fresnel equations are used to describe the reflection and transmission of the incident light. Due to a potential multiple reflection, absorption of the laser beam is also modified [83] and thereby a ray tracing method is employed to compute the absorbed power density while a similar methodology is ensued to estimate the proportion of the refraction by applying Snell's law. With respect to the numerical scheme used to simulate dynamics of velocity and pressure fields, a similar procedure is ensued in the event of subsequent pulses, however in this case the interaction with the modified surface profile induced by the first pulse due to the hydrodynamic motion of the molten material and its subsequent resolidification, should be taken into account. While second order finite difference schemes appear to be accurate for $NP$=1 where the surface profile has not been modified substantially, finer meshes and higher order methodologies are performed for more complex and profiles [84,85]. By taking into account the above, in corrugated surfaces Eq.7 instead of Eq.4 is used.

Regarding the boundary conditions for the fluid dynamics, the following constraints are imposed:

1. $\vec{u} = 0$, on the solid-liquid interface (non-slipping conditions),

2. $\mu \dfrac{\partial \vec{u}_t}{\partial z} = \dfrac{\partial \sigma}{\partial T_s} \dfrac{\partial T_s}{\partial r} \bigg|_{\vec{r} \text{ on surface}}$, on the top surface ($T_s$ is the surface temperature and $\sigma$ stands for the surface tension). Its gradient along the tangential direction $\vec{u}_t$ is balanced by shear stress (the position vector, $\vec{r}$, is taken on the surface of the molten profile),

3. $P$=0, on the top surface.

Numerical integration is allowed to move to the next time step provided that all variables at every element satisfy a predefined convergence tolerance of ±0.1%. To simplify the solution procedure, it is noted that hydrodynamic equations are solved in regions that contain either solid or molten material. To include the 'hydrodynamic' effect of the solid domain, material in the solid phase is modelled as an extremely viscous liquid ($\mu_{solid} = 10^5 \mu$), which will result into velocity fields that are infinitesimally small. An apparent viscosity is then defined with a smooth switch function similar [20] to emulate the step of viscosity at the melting temperature. Resolidification is considered when the lattice temperature of the molten material drops below $T_{melt}$. With respect to the material removal part of the simulation, lattice and carrier temperatures are computed over time and if lattice temperature reaches ~ $T_{rem}$, mass removal is assumed [20,82]. In that case, the associated nodes on the mesh are eliminated and new boundary conditions of the aforementioned form on the new surface are enforced. In order to preserve the smoothness of the surface that has been removed and allow an accurate and non-fluctuating value of the computed curvature and surface tension pressure, a fitting methodology is pursued.

## IV. EXPERIMENTAL PROTOCOL

TABLE I. Model parameters for GaAs.

| Solid Phase | | |
|---|---|---|
| **Quantity** | **Symbol (Units)** | **Value** |
| Initial temperature | $T_0$ (°K) | 300 |
| Electron-hole pair heat capacity [86] | $C_c$ (J/m$^3$ K) | $3N_e k_B$ |
| Lattice heat capacity [87] | $C_L$ (J/m$^3$ K) | $(3.235 \times 10^5 + 52.81\, T_L - 1.48 \times 10^9\, T_L^{-2})\, \rho_L$ |
| Lattice heat conductivity [88] | $K_L$ (W/m K) | $7.1 \times 10^4\, T_L^{-1.25}$ |
| Band gap energy [89] | $E_g$ (J) | $2.4337 \times 10^{-19} - 8.6646 \times 10^{-23}\, T_L^2/(T_L + 204)$ |
| Interband absorption (800nm) [90] | $\alpha_{SPA}$ (m$^{-1}$) | $2.91 \times 10^6\, e^{3.22(1.184 - E_g)}$ |
| Two-photon absorption (800nm) [91] | $\beta_{TPA}$ (sec m/J) | $1 \times 10^{-9}$ |
| Auger recombination coefficient [92] | $\gamma$ (m$^6$/sec) | $1.1 \times 10^{-43}$ |
| Impact ionisation coefficient [32] | $\theta$ (sec$^{-1}$) | $3.6 \times 10^{10}\, e^{-1.5 E_g / k_B T_c}$ |
| Carrier-phonon relaxation time [93] | $\tau_e$ (sec) | $\tau_{e0}\left[1 + \left(\dfrac{N_e}{N_{th}}\right)^2\right]$, $\tau_{e0}$ =0.5 ps, $N_{th}$=2 × 10$^{21}$ cm$^{-3}$ |
| **Molten Phase** (indicated by the *(m)* superscript) | | |
| Lattice heat capacity [94] | $C_L^{(m)}$ (J/m$^3$ K) | $0.46\, \rho_L$ |
| Lattice heat conductivity [95] | $K_L^{(m)}$ (W/m K) | $7.54 \times 10^4 \left(T_L^{(m)}\right)^{-1.29}$ |
| Density [95] | $\rho_L$ (gr/m$^3$) | $-8.2917 \times 10^{-3}\left(T_L^{(m)}\right)^2 - 85.624\left(T_L^{(m)}\right) + 5.3429 \times 10^6$ |
| Dynamic viscosity [96] | $\mu$ (gr/m sec) | $-3.717 \times 10^{-17}\left(T_L^{(m)}\right)^3 - 1.569 \times 10^{-3}\left(T_L^{(m)}\right)^2 - 2.208\left(T_L^{(m)}\right) + 1036$ |
| Surface tension [97] | $\sigma$ (N/m) | $0.401 - 0.18 \times 10^{-3}\, (T_L^{(m)} - T_{melt})$, for $T_L < 1610$ K [97]<br>$0.5821\left(T_L^{(m)}\right)^3 - 6.8871 \times 10^{-5}\left(T_L^{(m)}\right)^2 - 4.7281 \times 10^{-8}\, T_L^{(m)} + 7.5591 \times 10^{-12}$, for 1610 K $< T_L <$ 3 $T_{melt}$ [63,97] |
| Melting temperature | $T_{melt}$ (K) | 1511 |
| Latent heat of melting [95] | $L_{melt}$ (J/m$^3$) | $3.783 \times 10^9$ |

A Titanium Sapphire (Ti:Sa) regenerative amplifier laser was used with central wavelength at 800 nm, at 1 kHz repetition rate, of 100 fs pulse duration and 5 mJ pulse energy. The laser beam was guided through an optical microscope setup onto a semi-insulating Gallium Arsenide (SI-GaAs) substrate. A pyroelectric sensor was used to measure the power on the sample surface (~200 mJ/cm$^2$). The irradiation took place in room conditions at pressure equal to 101.325 kPa (1 atm). A two dimensional high precision translation stage was used to move the sample with spatial resolution of 1 µm. The sample was fixed on a two axis translation stage that was set to move at a constant velocity equal to 300 µm/s perpendicularly to the laser beam direction. By setting the repetition rate of the laser at 1 kHz and the beam spot size of 4.24 µm on the focal plane of the surface, each point of the material was irradiated with 14 pulses (on average). The beam moved relatively to the sample in a zig-zag motion scanning a square area of 50 µm × 50 µm. On the sample shown in Fig.2, the laser beam is scanning along the x- direction but making small steps on the y-direction. The y-step length is selected to match the ablation spot diameter so the overlap is minimal to none (in the y-direction) leaving no gaps between the ablation areas. Along the x-direction, the translation stage is moving at a constant velocity with a constant pulse repetition rate of 1KHz, causing overlap of each ablation spot. From shot to shot the overlap is calculated to 93%. The time scale of dynamics investigated in these experiments (>50 ps) is much shorter than the delay between consecutive pulses (1 ms for 1 KHz lase amplifier system). For every repetition, the fluence for each pulse has an isolated effect and is not influenced by the next pulse but instead the sum of pulses has an accumulative effect on the reshaping of the geometry of the surface.

A parametric analysis was performed by varying the scanning velocity, the beam overlapping area of ablation and fluence relative to the scanning direction. Since comparison of results as a function of scanning velocity is not convenient, the scanning velocity is converted to the number of pulses per spot through the relation with the laser repetition rate (hence only the *NP* is used in this work and not the scanning velocity). To characterize the produced surface profiles, a scanning electron microscope (SEM) was used (Fig.2a). A two dimensional Fourier transform (2DFT) was applied, to measure the spatial frequency of the ripples in a zoomed area of dashed box on the ablated sample (Fig.2b). It is illustrated that the ripples are orientated perpendicularly to the linear polarization of the laser beam for a linearly polarized (*p*-polarized) beam. In the 2DFT plot Fig.2c, three spatial frequencies are identified. The lowest frequencies around zero in the middle of the graph, correspond to patterns with low to no periodicity or noise. By contrast, the frequencies around $\pm 1.6$ μm$^{-1}$ correspond to periodic structures related to LIPSS. Finally, the highest frequencies are related with the side edges, two in each ripple, thus the double spatial frequency compared to the main body and image noise patterns ([63]). Those frequencies are not predicted by the SP model. The areas containing the spatial frequencies of the LIPSS are plotted by performing in inverse Fourier transform to identify the ripple formation position (Fig.2d). The 2DFT analysis and the bandwidth of the spatial frequencies of LIPSS yield a dispersion of the ripple periodicities in a range between 550 nm and 680 nm.

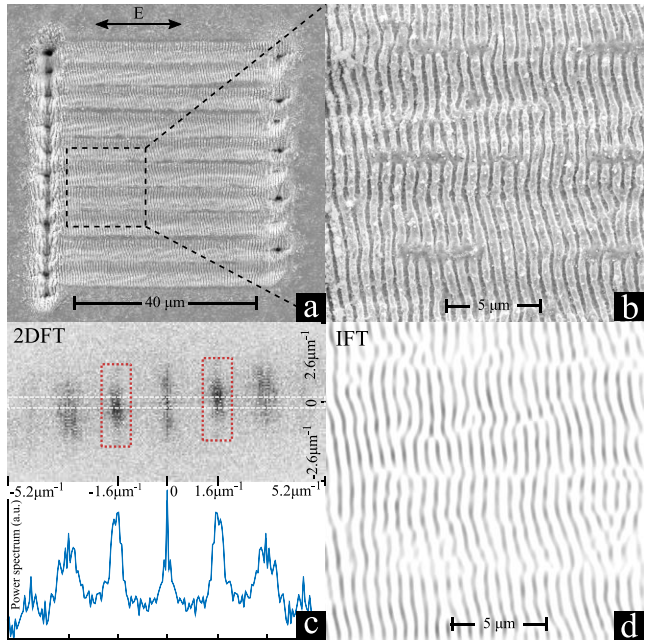

FIG. 2 (Color Online): SEM image of the laser ablated surface of GaAs sample and analysis: (a) complete view of the scanned area ($\vec{E}$ stands for the laser beam polarisation), (b) zoomed area of (a), (c) 2D Fourier transform of the area in (b), (d) inverse Fourier transform from the areas in (c) red boxes at $\pm 1.6$ μm$^{-1}$ generating a rippled region with periodicities in a range between 550 nm and 680nm. ($E_p$=200 mJ/cm$^2$, $\tau_p$=100 fs, *NP*=14).

## V. RESULTS AND DISCUSSION

### a. Ultrafast Dynamics and SP excitation

The ultrafast dynamics and the thermal response of the heated material are investigated to take into consideration the role of fluence and the pulse duration. Firstly, laser conditions for irradiation of GaAs with a single pulse of fluence $E_p = 70$ mJ/cm$^2$ and pulse duration $\tau_p = 100$ fs are assumed. The evolution of the carrier density and the carrier and lattice temperatures are illustrated in Fig.3. Similar to results for Si, the behaviour in the nonequilibrium regime is due to the difference between the temporal scales of the pulse duration and the electron-phonon relaxation times [20,32,36,59,60]. The carrier temperature remarkably increases during the first moments of irradiation (see inset in Fig.3) due to the fact that the heat capacity of the carrier system is several orders of magnitude larger than that of the lattice. The main processes of energy gain of the electron system through a (linear) one-photon and free carrier energy absorption. Following a short period when the carrier system does not further increase its energy, $T_c$ rises rapidly as the available pulse energy increases. To explain the initial increase followed by a slightly decreasing carrier temperature (*clamped region*) that occurs before the amount of the absorbed energy becomes significant which, subsequently, leads to a rapid rise of $T_e$, one has to examine the contribution of the competing mechanisms indicated by the various components in the '*source term*' (Eq.2). As demonstrated for Silicon [20,32,59,60], fluence plays a very significant role in the shape of the $T_e$ curve. More specifically, at lower fluences (Fig.2a in Ref [63]), there is initially, a small amount of electron-hole pairs that leads to a small carrier heat capacity. As a result, Eq.2 yields a noticeable increase of $T_e$ which continues until (i) the energy absorption is large enough to generate a sufficient amount of carriers, and (ii) the energy loss due to a pronounced thermal transfer from the carriers to the lattice. The increase of these two factors lead to a decrease of the thermal energy of the carriers that is also reflected from the decrease of $T_e$ (i.e. due to the increase of the relevant terms in Eq.2). On the other hand, larger fluences produce larger amounts of carriers and (given the fact that one photon-absorption and Auger recombination are the two predominant factors that alter carrier density), Auger recombination becomes significantly important as it varies as $N_e^3$. Hence, the enhanced Auger recombination (Fig.2b in Ref [63]) converts carrier ionisation energy into kinetic energy that results into an increase of the carrier temperature before it starts falling again. Therefore, for moderate values of fluence, a two peak structure of $T_e$ is shown. Finally, at even larger fluences (used in these simulations), the first peak almost disappears yielding a slow decrease (*clamped region*) before a much higher peak is produced that occurs after the carrier density has reached its peak (Fig.3). It is evident around the time the pulse yields the peak intensity, the carriers acquire their highest energy and in that case, two-photon and free carrier absorption processes have a significant influence. The increase of the carrier energy even after the time at which their density has reached its maximum is due to the free carrier absorption. After the carriers reach the highest energy, their temperature start to decrease mainly because of their interaction with the lattice while Auger recombination delays the heating of the lattice. On the other hand, the decrease of the carrier density is caused by the Auger recombination effect (Fig.3). It is noted that $T_e$ attains very high values and this could potentially influence the energy band gap response. Nevertheless, the adequate description of carrier dynamics and thermal response of Silicon [20,32,59,60] by ignoring a dramatic change in the band gap for large $T_e$ indicates that the influence of very energetic carriers is insignificant to $E_g$. Therefore, even for GaAs, a $T_L$-dependence is assumed regardless of the magnitude of $T_e$. Certainly, a more rigorous approach (e.g. based on first principles) should also be considered to estimate a possible $E_g$ variation for large $T_e$.

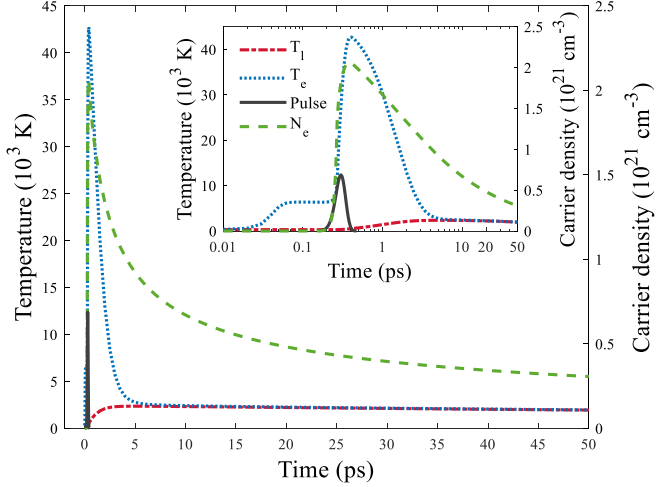

FIG. 3 (Color Online): Evolution of the carrier density, electron and lattice temperatures at $x=y=z=0$ ($E_p$=70 mJ/cm$^2$, $\tau_p$=100 fs).

The reflectivity of the irradiated material drops rapidly within the pulse duration to a minimum value that corresponds to $Re\ (\varepsilon)=1$ (Fig.4) (see also Ref. [65]) before it starts rising as the density of the excited carriers increases sharply. The Auger recombination related decrease of $N_e$ leads to a gradually relaxation of reflectivity towards the initial value of the unexcited material. On the other hand, Fig.4 shows that the density becomes large enough to satisfy the criterion for excitation of SP (i.e. $Re\ (\varepsilon) < -1$). Furthermore, the steep increase of the reflectivity around the 'metallisation' of the material (i.e. $Re\ (\varepsilon) = 0$) is demonstrated in Fig.4. The metallic behavior is also followed by a phase transition the material undergoes (i.e. melting).

To illustrate the effect of fluence on the optical properties of the irradiated material and the induced energy absorption that is expected also to influence the thermal response of the system (and equilibration process) as well its surface morphology, the evolution of reflectivity is investigated for various values of $E_p$. Simulation results indicate that there exists a faster acquisition of the minimum value (shift of the minimum to smaller times in Fig.5a) which is attributed to the generation of an increased number of carriers at higher fluences. Unlike the behavior for small values, at larger fluences the reflectivity increases abruptly (around $Re\ (\varepsilon) = 0$ as seen in Fig.5a demonstrating a metallic behaviour). The variation of fluence not only influences the temporal evolution of the reflectivity but also its absolute maximum change from the initial value at $t=0$, $\mathbb{R}$. It is shown (Fig.5b) that $\mathbb{R}$ which is taken to be around the moment of phase transition ($t$~350 fs) decreases for small values of the laser fluence while it increases at larger $E_p$ (where the metallic behaviour is more pronounced.

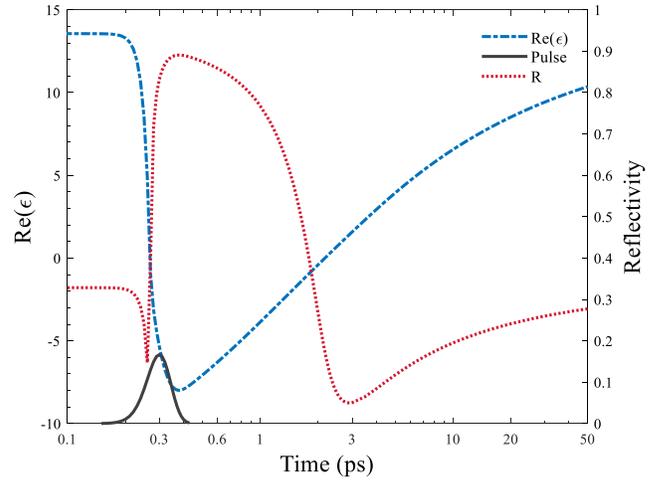

FIG. 4 (Color Online): Evolution of the real part of the dielectric constant at $x=y=z=0$. ($E_p$=70 mJ/cm$^2$, $\tau_p$=100 fs).

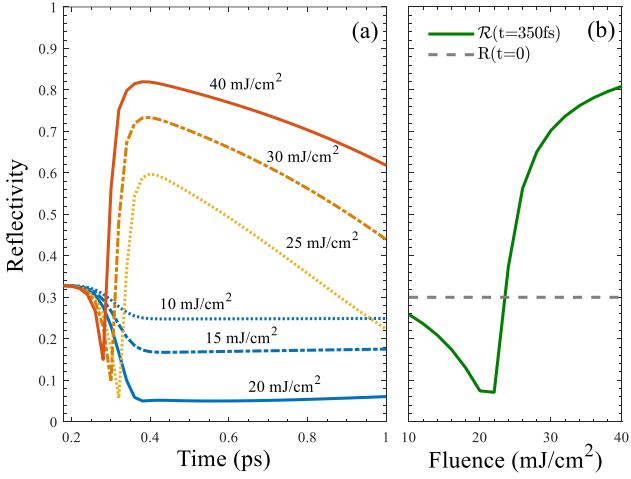
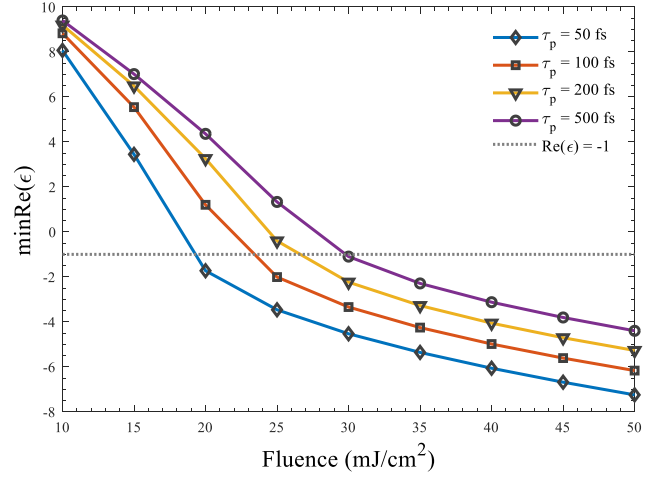

FIG. 5 (Color Online): (a) Reflectivity evolution at various fluences at $z=0$ ($\tau_p=100$ fs), (b) Absolute value of the difference between the maximum change of reflectivity from the initial value, $\mathbb{R}$, as a function of fluence.

FIG. 6 (Color Online): Dependence of the minimum value of the real part of the dielectric constant as a function of fluence for various values of the pulse duration.

On the other hand, the role of both the fluence and the pulse duration in the response of the dielectric constant is illustrated in Fig.6. More specifically, it is evident that the real part of the dielectric constant drops at increasing fluence for a specific $\tau_p$. To interpret the results, it is noted that when the semiconductor is in an unexcited state, the condition *Re (ε) >0* holds (despite the dielectric constant is dependent on the laser wavelength, this condition is always valid). As the carrier density increases, the material moves to the 'metallic' regime. Therefore, higher energies are expected to allow the resonance condition (i.e. Re (ε) =0) to be met easier as unlike a small absorption efficiency at small intensities, energy absorption is more pronounced due to a significant contribution of multi-photon-assisted excitation [36]. Similarly, keeping the fluence constant, a higher energy absorption is achieved when the temporal width of the pulse is smaller which results in an enhanced density of excited carriers. In Fig.6, it is shown that the laser conditions suffice to excite SP (*Re (ε) <-1*).

To emphasise on the features of the excited SP, a spatio-temporal representation of the real part of the dielectric constant (Fig.7) is illustrated. The *white* line defines the boundaries of the region in which the material is characterized by *Re (ε) <-1* which allows the excitation of SP. According to the illustrated picture, the maximum damping depth is of the order of 18 nm if the SP conditions are satisfied. This is similar to the simulated value for Si [36].

On the other hand, the lifetime of the SP (i.e. smaller than the time required for melting and larger than the time needed to couple with the incident beam) as well the carrier density value (~2.6 × $10^{21}$ cm$^{-3}$ at around 80% of the intensity (Fig.8)) indicate that a SP-related periodic modulation of a resultant energy deposition through the coupling with the incident beam is possible (for this part, simulations have been performed for $E_p$=200 mJ/cm$^2$, $\tau_p$=100 fs to compare with experimental results). According to the computed value of the SP wavelength (~772 nm) based on the relevant $N_e$ (2.6 × $10^{21}$ cm$^{-3}$) and the use of Eq.9, there is a deviation from the experimental result (in the range of between 550 nm and 680 nm for *NP* = 14). This is an indication that the employment of Eq.9 towards calculating the SP-resonance wavelength is problematic as the expression was derived on the assumption that the surface morphology is flat (i.e. requirement of continuity of electric and magnetic field on flat profiles). Therefore, as suggested in Section II, a different methodology is required when *NP* is large enough to lead to the production of corrugated surfaces.

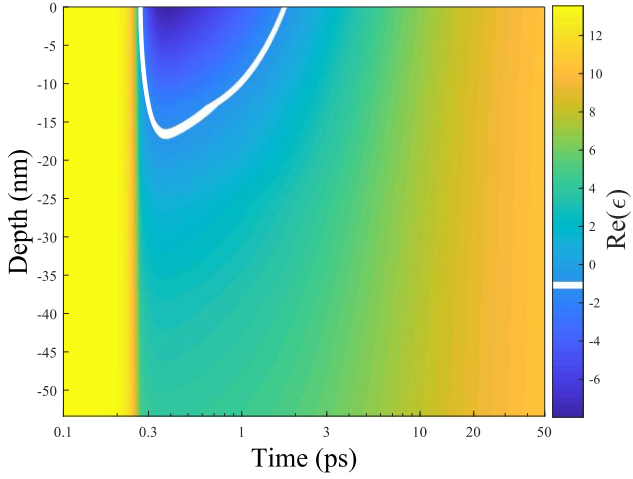

FIG. 7 (Color Online): Spatio-temporal evolution of the real part of the dielectric constant. *White* line define the limit where *Re (ε) <-1*. ($E_p$=70 mJ/cm$^2$, $\tau_p$=100 fs).

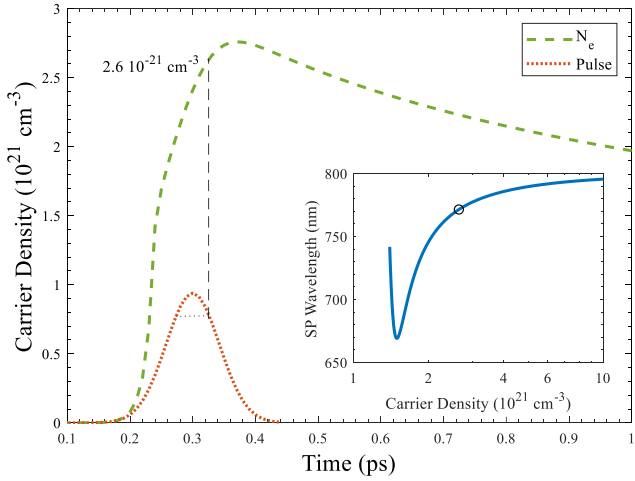

FIG. 8 (Color Online): Carrier density evolution as a function of time. The temporal intensity profile defines the carrier density that gives a resonance (at 80%). ($E_p$=200 mJ/cm$^2$, $\tau_p$=100 fs).

### b. Periodic structures and damage thresholds

In previous reports, a correlation of the SP wavelength on the corrugation depth was proposed based on Finite Difference Time Domain (FDTD) simulations [17,73]. While the evaluation of the optimum coupling of the laser beam with the excited SP requires appropriate calculations for GaAs, an estimate of the SP wavelength values can be deduced given the resemblance of the produced depths to those in our simulations. The dependence of the SP-wavelength $\Lambda_{SP}$ as a function of the corrugation height is illustrated in Fig. 9a (a blue shift to smaller wavelengths with increase of corrugation height as shown for Si [17]) while the approach to describe the multiscale physical mechanisms presented in Section II yield the correlation of the periodicity of the induced structures as a function of the number of pulses (Fig. 9b). To correlate the number of pulses with the induced ripple's corrugation features (bottom-to-peak distance), Eqs.1-14 were solved together to calculate the effect of ablation and hydrodynamical transport impact; the latter is illustrated through the analysis of the fluid transport that is due to the temperature gradient and induced Marangoni mechanism of mass transfer [20,63]). The surface profile produced from the simulations is illustrated for *NP* = 14 in a sector of the heat-affected region (Fig.10) where the rippled profile is characterized by an average periodicity *Λ* that is approximately equal to 637 nm. It is shown that the orientation of the induced periodic structures are perpendicular to the polarization of the electric field of the incident beam (indicated by the [32] *black doubled arrow* in Fig.10). The simulated result for $NP = 14$ for the ripple frequency (~ 637 nm while the SP wavelength is 648 nm) appears to be close to the experimental result (Fig.2). The value difference between the SP wavelength and the predicted result is due to the correction assuming the role of fluid dynamics [20,39]. Certainly, conditions that could improve the agreement with experimental observations include the consideration of a more precise mass removal mechanism of the material due to intense heating [20,39], as a deeper profile is expected (Fig.9a) to lead to a larger shift of SP wavelength to smaller values. An expected damage threshold sharp decrease due to incubation effects (i.e. formation of self-trapped excitons) could also lead to an increase of the profile depth that is necessary to induce the deeper profile [98,99]. However, as a detailed description of the role of incubation effects for GaAs is still elusive [98,99], further analysis especially at high temperatures cannot be performed in a conclusive and accurate way. One approach to attain a more accurate description is to incorporate molecular dynamics into the continuum modelling methodology [100].

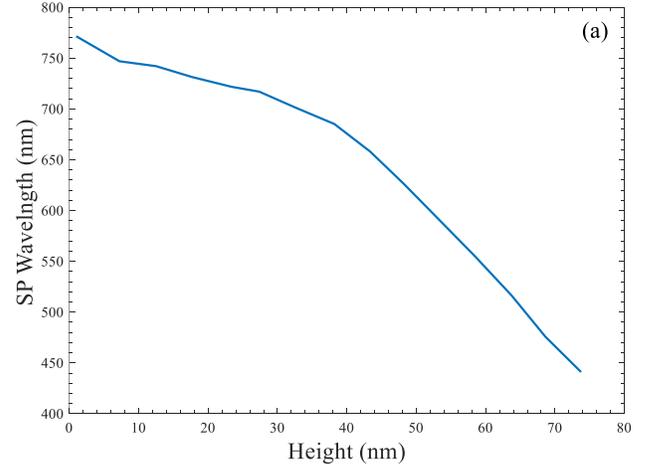

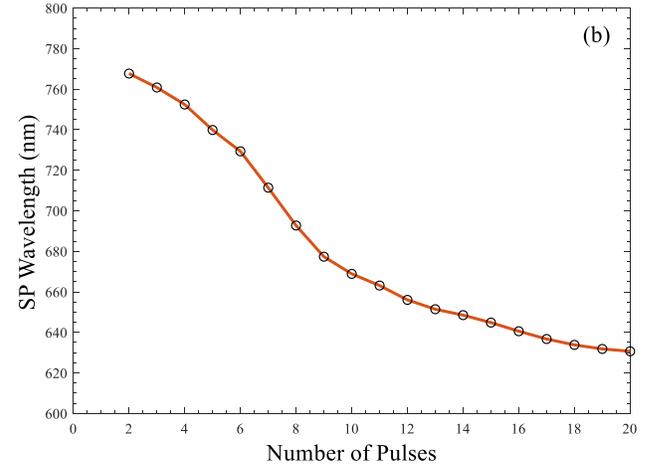

FIG. 9 (Color Online): (a) SP wavelength *vs.* corrugation height, (b) SP periodicity *vs. NP* ($E_p$=200 mJ/cm$^2$, $\tau_p$=100 fs).

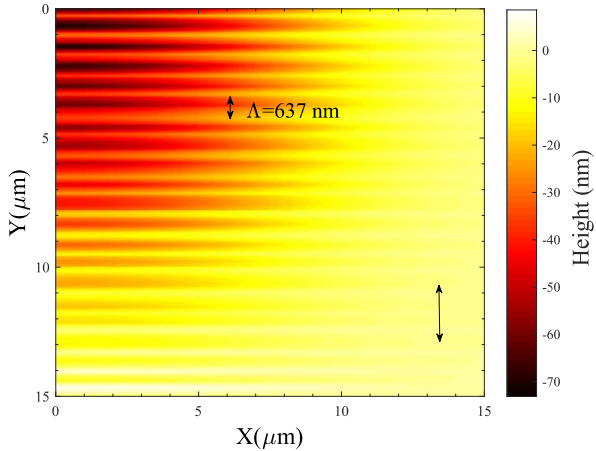

FIG. 10 (Color Online): Surface pattern in a quadrant. *Black doubled arrow* at the bottom right indicates the polarization direction of the electric field of the incident beam. $\Lambda$ stands for the calculated rippled periodicity. ($NP=14$, $E_p=200$ mJ/cm$^2$, $\tau_p=100$ fs).

The decrease of the SP wavelength with increasing dose (i.e. $NP$) that is reflected on the periodicity of the rippled profile (Fig.9b) follows the pattern also observed or predicted in other materials (see [14,20,21,23-25,27,31,37] and references therein). More specifically, a steep decrease is followed by saturation point. Although it is of paramount importance to explore whether (and at what values of $NP$), different periodic structures (i.e. grooves or spikes) are formed [11,31,101], such an investigation is beyond the scope of the present study.

While a detailed description of the physical mechanisms that lead to LIPSS formation is expected to allow production of the periodic structure formation in a systematic way, another significant area of investigation is the correlation of the damage thresholds (for $NP=1$) with the laser parameters and more specifically the laser pulse duration. In regard to the determination of the value of the 'damage threshold', it is noted that there is an ambiguity regarding to whether morphological damage is associated to a mass displacement due to melting and resolidification of the material or it is related to a mass removal (i.e. or ablation). In Fig. 11, the computed fluence ('damage threshold') corresponds to the minimum fluence value that induces melting of the material and it is plotted against the pulse duration. As explained previously, a pulse width increase leads to a decrease of the absorbed energy (reflected also on the reduced number of excitation carriers shown in Fig. 11) which indicates that more energy is required to damage the surface. Hence, in that case, the pulse should be characterised by a higher intensity that increases the damage threshold. Simulation results show that the damage threshold varies as $\sim \tau_p^{\gamma}$ at small pulse durations (<1 ps, where $\gamma = 0.053\pm0.011$). Similar dependencies have been estimated for other materials [51,59,90]. By contrast, at larger $\tau_p$, there is a pronounced decrease of the damage threshold which is explained by an increased maximum lattice temperature [59]; the decline of the damage threshold at increasing pulse durations (larger than a few ps) has also been reported in previous studies [32,59,90]. The theoretical prediction is tested against experimental data in previous reports, nevertheless, those experimental results correspond to ablation studies or in other conditions [33,54-56]. More specifically, in a previous report, $E_p^{\text{damage}}$ = 100 mJ/cm² for $\tau_p$ = 70 fs while experimental results in this work and an analysis of the SEM images (Fig.2) show that the laser conditions ($E_p^{\text{damage}}$ = 200 mJ/cm² for $\tau_p$ = 100 fs) produce some small mass removal. Due to the fact that mass removal requires large temperatures which in other materials is estimated to be about 2-3 times larger than the melting point (considering temperatures above the boiling or critical points [20,25]) and given the, approximately, linear dependence of the temperature from the fluence, a rough estimate can be deduced to project the damage thresholds in Fig. 11. Therefore, an estimated 35-37 mJ/cm² is deduced for a (experimental) value for the melting threshold for $\tau_p$ = 100 fs which is in a good agreement with the theoretical predictions (Fig.11). Nevertheless, a more accurate conclusion will be drawn if more appropriately developed experimental (for example, time-resolved experimental) protocols are introduced to evaluate the damage thresholds at the onset of the phase transition. To the best of our knowledge, there are not similar reports with experimental results for the pulse durations explored in this study.

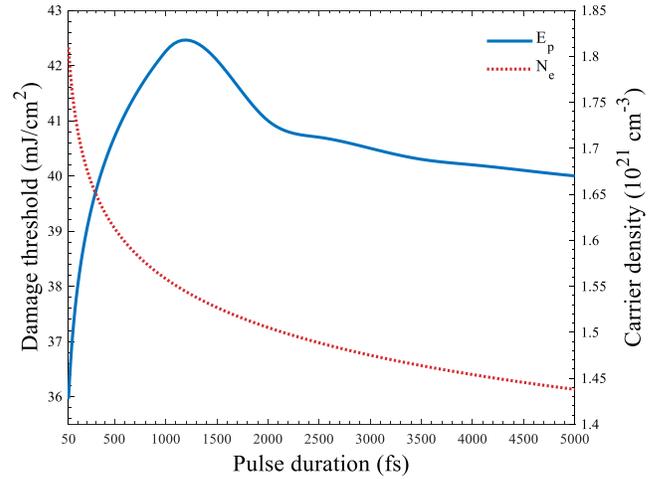

FIG. 11 (Color Online): Damage thresholds and carrier density as a function of the pulse duration at $z=0$.

Furthermore, apart from the estimation of the damage threshold and the frequencies of the periodic structures, the methodology can be used to derive further a complete description of the evolution of the morphological features of the induced structures (such as height, depth, volume of ablated region, etc.) as well as an elaborated account of the fluid dynamics (see [63]). Similar approaches have been ensued in previous reports [20,24], however, a detailed presentation of such results is beyond the scope of the current work.

The theoretical model presented in this work aimed to describe systematically the parameters that influence the ultrafast dynamics of GaAs after heating with ultrashort pulses and estimate the thermal response of the material (associated eventually with surface modification); nevertheless, there are still many unexplored issues that need to be addressed (i.e. effective mass dependence on fluence, wavelength dependence, excitation in very short pulses, role of more complex beam polarisation states, structural effects in extreme conditions, distinction between amorphous or crystalline material, mechanical effects, more accurate behaviour in ablation conditions, formation of voids inside the material after repetitive irradiation, role of incubation effects, consideration of degeneracy and departure from a Maxell-Boltzmann-based assumption of carrier distribution, precise computation of carrier collision frequency and electron-phonon coupling, inclusion of the influence of impact ionisation and carrier diffusion, etc.) before a complete picture of the physical processes that characterise heating of GaAs with femtosecond laser pulses is attained.

## VI. CONCLUSIONS

A detailed theoretical framework was presented that describes, for the first time, both the ultrafast dynamics and surface modification physical mechanism after heating of GaAs with ultrashort pulsed lasers. The influence of the laser conditions such as the pulse duration and the fluence were evaluated in an effort to explore conditions that lead to SP excitation. Results show an increase of the maximum value of the reflectivity at increasing fluence that emphasises the metallic character of the heated material. To account for the frequencies of the experimentally observed periodic structures, a dose-dependent modulation of the SP-wavelength which is excited on the corrugated surface is performed. Simulation results revealed a $\sim \tau_p^{0.053\pm0.011}$ (at relatively small values of the pulse duration) dependence of the damage threshold and a SP-related periodic structure formation. The correlation of the laser characteristics dependencies of the ultrafast dynamics, material damage and onset of periodic structure formation can be used to streamline the modulation of the frequencies of the structures on the surface of a still not fully explored semiconductor under intense heating. Hence, a detailed description of the thermal response of the material is aimed to allow a systematic laser-based processing and produce surface structures with application-based properties.

## ACKNOWLEDGEMENTS

G.D.T acknowledges financial support from *Nanoscience Foundries and Fine Analysis (NFFA)–Europe* H2020-INFRAIA-2014-2015 (under Grant agreement No 654360).

*Corresponding author: amargiol@physics.uoc.gr

[1] A. Y. Vorobyev and C. Guo, Laser & Photonics Reviews **7**, 385 (2012).
[2] V. Zorba, L. Persano, D. Pisignano, A. Athanassiou, E. Stratakis, R. Cingolani, P. Tzanetakis, and C. Fotakis, Nanotechnology **17**, 3234 (2006).
[3] V. Zorba, E. Stratakis, M. Barberoglou, E. Spanakis, P. Tzanetakis, S. H. Anastasiadis, and C. Fotakis, Advanced Materials **20**, 4049 (2008).
[4] D. Bäuerle, *Laser processing and chemistry* (Springer, Berlin; New York, 2000), 3rd rev. enlarged edn.
[5] J.-C. Diels and W. Rudolph, *Ultrashort laser pulse phenomena : fundamentals, techniques, and applications on a femtosecond time scale* (Elsevier / Academic Press, Amsterdam ; Boston, 2006), 2nd edn.
[6] E. L. Papadopoulou, A. Samara, M. Barberoglou, A. Manousaki, S. N. Pagakis, E. Anastasiadou, C. Fotakis, and E. Stratakis, Tissue Eng Part C-Me **16**, 497 (2010).
[7] Z. B. Wang, M. H. Hong, Y. F. Lu, D. J. Wu, B. Lan, and T. C. Chong, Journal of Applied Physics **93**, 6375 (2003).
[8] R. Böhme, S. Pissadakis, D. Ruthe, and K. Zimmer, Applied Physics A-Materials Science & Processing **85**, 75 (2006).
[9] S. M. Petrovic, B. Gakovic, D. Perusko, E. Stratakis, I. Bogdanovic-Radovic, M. Cekada, C. Fotakis, and B. Jelenkovic, Journal of Applied Physics **114**, 233108 (2013).
[10] B. N. Chichkov, C. Momma, S. Nolte, F. vonAlvensleben, and A. Tunnermann, Applied Physics a-Materials Science & Processing **63**, 109 (1996).
[11] J. J. J. Nivas, S. He, A. Rubano, D. Paparo, L. Marrucci, R. Bruzzese, and S. Amoruso, Sci Rep-Uk **5**, 17929 (2015).
[12] M. Beresna, M. Gecevicius, P. G. Kazansky, and T. Gertus, Applied Physics Letters **98**, 201101 (2011).
[13] E. Skoulas, A. Manousaki, C. Fotakis, and E. Stratakis, Sci Rep-Uk **7**, 45114 (2017).
[14] G. D. Tsibidis, E. Skoulas, and E. Stratakis, Optics Letters **40**, 5172 (2015).
[15] E. Stratakis, A. Ranella, and C. Fotakis, Biomicrofluidics **5**, 013411 (2011).
[16] J. Bonse, M. Munz, and H. Sturm, Journal of Applied Physics **97**, 013538 (2005).
[17] M. Huang, F. L. Zhao, Y. Cheng, N. S. Xu, and Z. Z. Xu, ACS Nano **3**, 4062 (2009).
[18] J. E. Sipe, J. F. Young, J. S. Preston, and H. M. Vandriel, Physical Review B **27**, 1141 (1983).
[19] Z. Guosheng, P. M. Fauchet, and A. E. Siegman, Physical Review B **26**, 5366 (1982).
[20] G. D. Tsibidis, M. Barberoglou, P. A. Loukakos, E. Stratakis, and C. Fotakis, Physical Review B **86**, 115316 (2012).
[21] G. D. Tsibidis, E. Skoulas, A. Papadopoulos, and E. Stratakis, Physical Review B **94**, 081305(R) (2016).
[22] G. D. Tsibidis, E. Stratakis, and K. E. Aifantis, Journal of Applied Physics **112**, 089901 (2012).
[23] A. Papadopoulos, E. Skoulas, G. D. Tsibidis, and E. Stratakis, Applied Physics A **124**, 146 (2018).
[24] G. D. Tsibidis and E. Stratakis, Journal of Applied Physics **121**, 163106 (2017).
[25] G. D. Tsibidis, A. Mimidis, E. Skoulas, S. V. Kirner, J. Krüger, J. Bonse, and E. Stratakis, Applied Physics A **124**, 27 (2017).
[26] M. Birnbaum, Journal of Applied Physics **36**, 3688 (1965).
[27] J. Bonse, S. Höhm, S. V. Kirner, A. Rosenfeld, and J. Krüger, Ieee J Sel Top Quant **23**, 9000615 (2017).
[28] Y. H. Han and S. L. Qu, Chemical Physics Letters **495**, 241 (2010).
[29] T. H. Her, R. J. Finlay, C. Wu, S. Deliwala, and E. Mazur, Applied Physics Letters **73**, 1673 (1998).
[30] A. J. Pedraza, J. D. Fowlkes, and Y. F. Guan, Applied Physics a-Materials Science & Processing **77**, 277 (2003).
[31] G. D. Tsibidis, C. Fotakis, and E. Stratakis, Physical Review B **92**, 041405(R) (2015).
[32] H. M. Vandriel, Physical Review B **35**, 8166 (1987).
[33] S. K. Sundaram and E. Mazur, Nature Materials **1**, 217 (2002).
[34] E. Knoesel, A. Hotzel, and M. Wolf, Physical Review B **57**, 12812 (1998).
[35] T. J. Y. Derrien, J. Kruger, T. E. Itina, S. Hohm, A. Rosenfeld, and J. Bonse, Applied Physics a-Materials Science & Processing **117**, 77 (2014).

[36]	T. J. Y. Derrien, T. E. Itina, R. Torres, T. Sarnet, and M. Sentis, Journal of Applied Physics **114**, 083104 (2013).
[37]	J. Bonse, A. Rosenfeld, and J. Krüger, Journal of Applied Physics **106**, 104910 (2009).
[38]	T. J. Y. Derrien, T. E. Itina, R. Torres, T. Sarnet, and M. Sentis, Journal of Applied Physics **114**, 083104 (2013).
[39]	M. Barberoglou, G. D. Tsibidis, D. Gray, E. Magoulakis, C. Fotakis, E. Stratakis, and P. A. Loukakos, Applied Physics A: Materials Science and Processing **113**, 273 (2013).
[40]	G. D. Tsibidis, E. Stratakis, P. A. Loukakos, and C. Fotakis, Applied Physics A **114**, 57 (2014).
[41]	O. Varlamova, F. Costache, J. Reif, and M. Bestehorn, Applied Surface Science **252**, 4702 (2006).
[42]	J. Z. P. Skolski, G. R. B. E. Romer, J. V. Obona, V. Ocelik, A. J. H. in 't Veld, and J. T. M. De Hosson, Physical Review B **85**, 075320 (2012).
[43]	J. Bonse, J. Krüger, S. Höhm, and A. Rosenfeld, Journal of Laser Applications **24**, 042006 (2012).
[44]	F. Garrelie, J. P. Colombier, F. Pigeon, S. Tonchev, N. Faure, M. Bounhalli, S. Reynaud, and O. Parriaux, Optics Express **19**, 9035 (2011).
[45]	J. Wang and C. Guo, Applied Physics Letters **87**, 251914 (2005).
[46]	J. Bonse and J. Kruger, Journal of Applied Physics **108**, 034903 (2010).
[47]	Y. Shimotsuma, P. G. Kazansky, J. R. Qiu, and K. Hirao, Physical Review Letters **91**, 247405 (2003).
[48]	B. Chimier, O. Utéza, N. Sanner, M. Sentis, T. Itina, P. Lassonde, F. Légaré, F. Vidal, and J. C. Kieffer, Physical Review B **84**, 094104 (2011).
[49]	G. D. Tsibidis, Applied Physics A **124**, 311 (2018).
[50]	G. D. Tsibidis, Journal of Applied Physics **123**, 085903 (2018).
[51]	C. S. R. Nathala, A. Ajami, W. Husinsky, B. Farooq, S. I. Kudryashov, A. Daskalova, I. Bliznakova, and A. Assion, Applied Physics A **122**, 107 (2016).
[52]	A. A. Ionin, S. I. Kudryashov, S. V. Makarov, L. V. Seleznev, and D. V. Sinitsyn, Applied Physics A **117**, 1757 (2014).
[53]	A. Borowiec and H. K. Haugen, Applied Physics Letters **82**, 4462 (2003).
[54]	A. M. T. Kim, J. P. Callan, C. A. D. Roeser, and E. Mazur, Physical Review B **66**, 5203 (2002).
[55]	J. P. Callan, A. M. T. Kim, C. A. D. Roeser, and E. Mazur, Ultrafast Physical Processes in Semiconductors **67**, 151 (2001).
[56]	K. Sokolowski-Tinten, H. Schulz, J. Bialkowski, and D. von der Linde, Applied Physics A **53**, 227 (1991).
[57]	S. Moon, K. Kim, Y. Kim, J. Heo, and J. Lee, Sci Rep-Uk **6**, 30107 (2016).
[58]	J. Madeo, A. Margiolakis, Z. Y. Zhao, P. J. Hale, M. K. L. Man, Q. Z. Zhao, W. Peng, W. Z. Shi, and K. M. Dani, Optics Letters **40**, 3388 (2015).
[59]	J. K. Chen, D. Y. Tzou, and J. E. Beraun, International Journal of Heat and Mass Transfer **48**, 501 (2005).
[60]	A. Rämer, O. Osmani, and B. Rethfeld, Journal of Applied Physics **116**, 053508 (2014).
[61]	S. I. Anisimov, Kapeliov.Bl, and T. L. Perelman, Zhurnal Eksperimentalnoi Teor. Fiz. **66**, 776 (1974 [Sov. Phys. Tech. Phys. 11, 945 (1967)]).
[62]	http://www.ioffe.ru/SVA/NSM/Semicond/GaAs/electric.html.
[63]	See Supplementary Material at [URL] for a detailed description of (a) electron temperature and carrier density evolution for different fluences, (b) approximate estimation of surface tension for diffferent lattice temperature values, (c) heat conductivity of carriers, (d) height of ripples at various *NP*, (e) spatial temperature for *NP*=14, (f) conditions for ablation.
[64]	H. M. van Driel, Physical Review B **35**, 8166 (1987).
[65]	K. Sokolowski-Tinten and D. von der Linde, Physical Review B **61**, 2643 (2000).
[66]	J. W. Pan, J. L. Shieh, J. H. Gau, J. I. Chyi, J. C. Lee, and K. J. Ling, Journal of Applied Physics **78**, 442 (1995).
[67]	W. Nakwaski, Physica B **210**, 1 (1995).
[68]	H. Raether, *Surface plasmons on smooth and rough surfaces and on gratings* (Springer-Verlag, Berlin ; New York, 1988), Springer Tracts in Modern Physics, 111.
[69]	J. M. Pitarke, V. M. Silkin, E. V. Chulkov, and P. M. Echenique, Reports on Progress in Physics **70**, 1 (2007).
[70]	A. M. Bonchbruevich, M. N. Libenson, V. S. Makin, and V. V. Trubaev, Opt Eng **31**, 718 (1992).
[71]	J. C. Wang and C. L. Guo, Journal of Applied Physics **102**, 053522 (2007).

[72]     C. A. Zuhlke, G. D. Tsibidis, T. Anderson, E. Stratakis, G. Gogos, and D. R. Alexander, AIP Advances **8**, 015212 (2018).
[73]     J. P. Colombier, F. Garrelie, P. Brunet, A. Bruyere, F. Pigeon, R. Stoian, and O. Parriaux, Journal of Laser Micro Nanoengineering **7**, 362 (2012).
[74]     J. E. Sipe, J. F. Young, J. S. Preston, and H. M. van Driel, Physical Review B **27**, 1141 (1983).
[75]     H. Y. Zhao, W. C. Niu, B. Zhang, Y. P. Lei, M. Kodama, and T. Ishide, Journal of Physics D-Applied Physics **44**, 485302 (2011).
[76]     J. H. Cho, D. F. Farson, J. O. Milewski, and K. J. Hollis, Journal of Physics D-Applied Physics **42**, 17, 175502 (2009).
[77]     D. P. Korfiatis, K. A. T. Thoma, and J. C. Vardaxoglou, Applied Surface Science **255**, 7605 (2009).
[78]     L. D. Landau and E. M. Lifshitz, *Fluid Mechanics* (2nd Edition) (1987).
[79]     J. Zhou, H. L. Tsai, and P. C. Wang, Journal of Heat Transfer-Transactions of the Asme **128**, 680 (2006).
[80]     R. Kelly and A. Miotello, Applied Surface Science **96-98**, 205 (1996).
[81]     N. M. Bulgakova and I. M. Bourakov, Applied Surface Science **197**, 41 (2002).
[82]     J. K. Chen and J. E. Beraun, J Opt a-Pure Appl Op **5**, 168 (2003).
[83]     P. Solana and G. Negro, Journal of Physics D-Applied Physics **30**, 3216 (1997).
[84]     Y. Morinishi, O. V. Vasilyev, and T. Ogi, Journal of Computational Physics **197**, 686 (2004).
[85]     Y. Morinishi, T. S. Lund, O. V. Vasilyev, and P. Moin, Journal of Computational Physics **143**, 90 (1998).
[86]     B. Schumann, Cryst Res Technol **26**, 18 (1991).
[87]     V. M. Glazov and A. S. Pashinkin, Inorg Mater+ **36**, 225 (2000).
[88]     R. Bogaard and A. N. Gerritsen, Phys Rev B-Solid St **3**, 1808 (1971).
[89]     J. S. Blakemore, Journal of Applied Physics **53**, R123 (1982).
[90]     J. R. Meyer, M. R. Kruer, and F. J. Bartoli, Journal of Applied Physics **51**, 5513 (1980).
[91]     F. Kadlec, H. Nemec, and P. Kuzel, Physical Review B **70**, 125205 (2004).
[92]     D. Steiauf, E. Kioupakis, and C. G. Van de Walle, ACS Photonics **1**, 643 (2014).
[93]     D. Agassi, Journal of Applied Physics **55**, 4376 (1984).
[94]     K. Itagaki and K. Yamaguchi, Thermochim Acta **163**, 1 (1990).
[95]     T. Kim, M. R. Pillai, M. J. Aziz, M. A. Scarpulla, O. D. Dubon, K. M. Yu, J. W. Beeman, and M. C. Ridgway, Journal of Applied Physics **108**, 013508 (2010).
[96]     K. Kakimoto and T. Hibiya, Applied Physics Letters **50**, 1249 (1987).
[97]     R. Rupp and G. Muller, Journal of Crystal Growth **113**, 131 (1991).
[98]     S. W. Winkler, I. M. Burakov, R. Stoian, N. M. Bulgakova, A. Husakou, A. Mermillod-Blondin, A. Rosenfeld, D. Ashkenasi, and I. V. Hertel, Applied Physics a-Materials Science & Processing **84**, 413 (2006).
[99]     A. Rosenfeld, M. Lorenz, R. Stoian, and D. Ashkenasi, Applied Physics a-Materials Science & Processing **69**, S373 (1999).
[100]   D. S. Ivanov and L. V. Zhigilei, Physical Review B **68**, 064114 (2003).
[101]   S. He, J. J. J. Nivas, K. K. Anoop, A. Vecchione, M. Hu, R. Bruzzese, and S. Amoruso, Applied Surface Science **353**, 1214 (2015).

# *Supplementary Material for*

**Ultrafast dynamics and sub-wavelength periodic structure formation following irradiation of GaAs with femtosecond laser pulses**

A. Margiolakis,[1,2] [*] G. D. Tsibidis,[3] K. M. Dani,[2] and G. P. Tsironis[1,3]

[1]*Department of Physics, University of Crete, P. O. Box 2208, 71003 Heraklion, Greece*

[2]*Femtosecond Spectroscopy Unit, Okinawa Institute of Science and Technology Graduate University, 1919-1 Tancha, Onna-son, Kunigami, Okinawa904-495, Japan*

[3] *Institute of Electronic Structure and Laser (IESL), Foundation for Research and Technology (FORTH), N. Plastira 100, Vassilika Vouton, 70013, Heraklion, Crete, Greece*

## I. Surface Tension dependence on lattice temperature

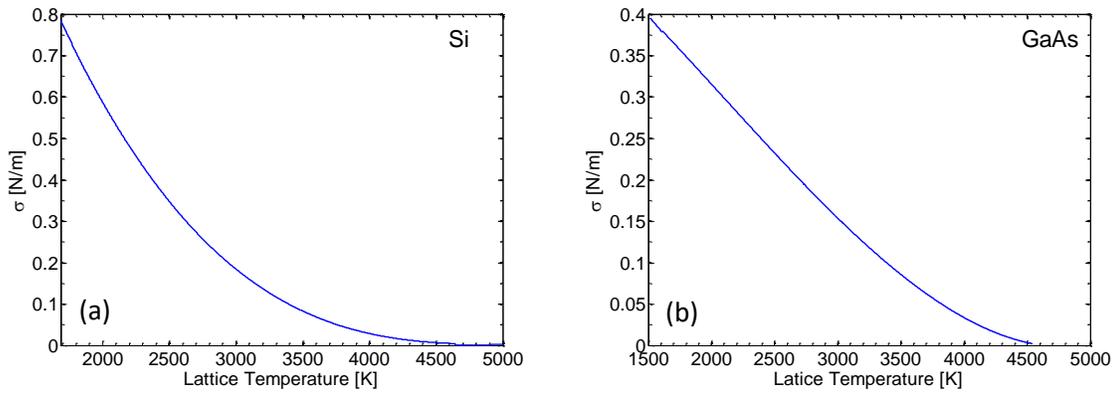

FIG. 1: Surface tension as a function of the lattice temperature for (a) Si and (b) GaAs.

As shown also in R. Rupp and G. Muller, *Journal of Crystal Growth* **113**, 131 (1991), the surface tension is temperature dependent. The expression that was used in our computations are $\sigma=0.401-0.18 \times 10^{-3} (T_L-T_m)$ ($T_L <1611$ K). To justify the latter value, we considered the evolution of the surface tension as a function of the temperature in the case of Si (G. D. Tsibidis, M. Barberoglou, P. A. Loukakos, E. Stratakis, and C. Fotakis, *Physical Review B* **86**, 115316 (2012)) (Fig.1a). More specifically, while at relatively small $T_L$ around the melting point, surface tension drops in a linear form, at larger temperatures the function becomes nonlinear while at relatively large values it vanishes. Similarly, an approximating function with a similar behavior (as in Si) for $\sigma$ for larger $T_L$ is produced and illustrated in Fig.1b.

## II. Carrier density and Electron temperature variation for various fluences

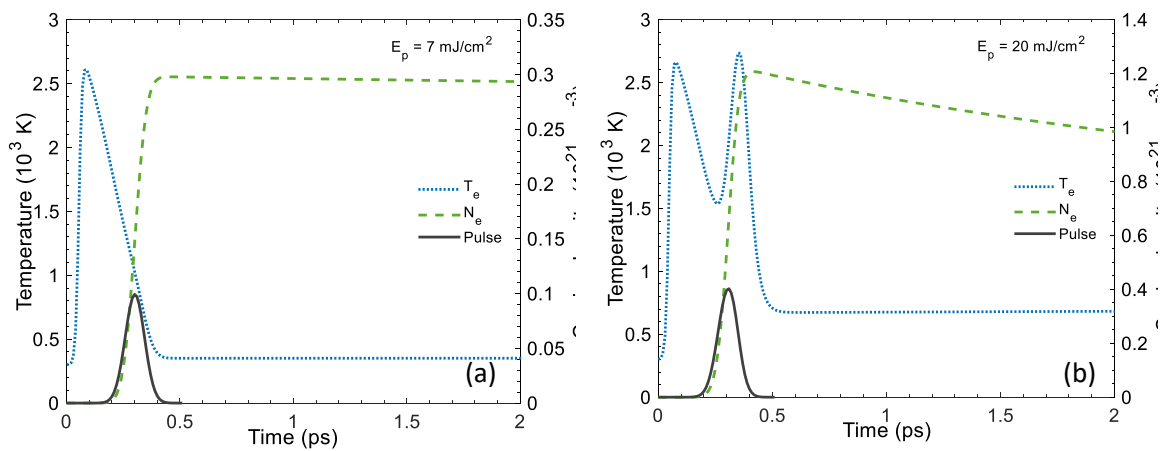

FIG. 2: Evolution of the carrier density and electron and lattice temperatures at $x=y=z=0$ for (a) $E_p$=7 mJ/cm$^2$ and (b) $E_p$=20 mJ/cm$^2$ ($\tau_p$=100 fs). The *black* curve indicates the temporal shape of the pulse.

## III. SEM IMAGE AND 2DFT ANALYSIS

In this Fig.3, a 2DFT was applied as in Figure 2 in the main manuscript. However, instead of plotting the spatial frequency at $\pm 1.6$ μm$^{-1}$, an inverse Fourier transform was applied at ~ $\pm 3.2$ μm$^{-1}$ to verify that the area with double the frequency of the ripples belongs to the edges of the ripples. More specifically, in (a) the area where 2DFT is applied is illustrated in (b) is the Fourier transform and the power spectrum taken from the area in the white dashed box. In (c) is the inverse Fourier transform from the red dashed box in (b). In (d) we have overlapped (c) in red color on top of (a), to make clear the distinction of the points that define spatial frequencies of the 2DFT are at ~ $\pm 3.2$ μm$^{-1}$.

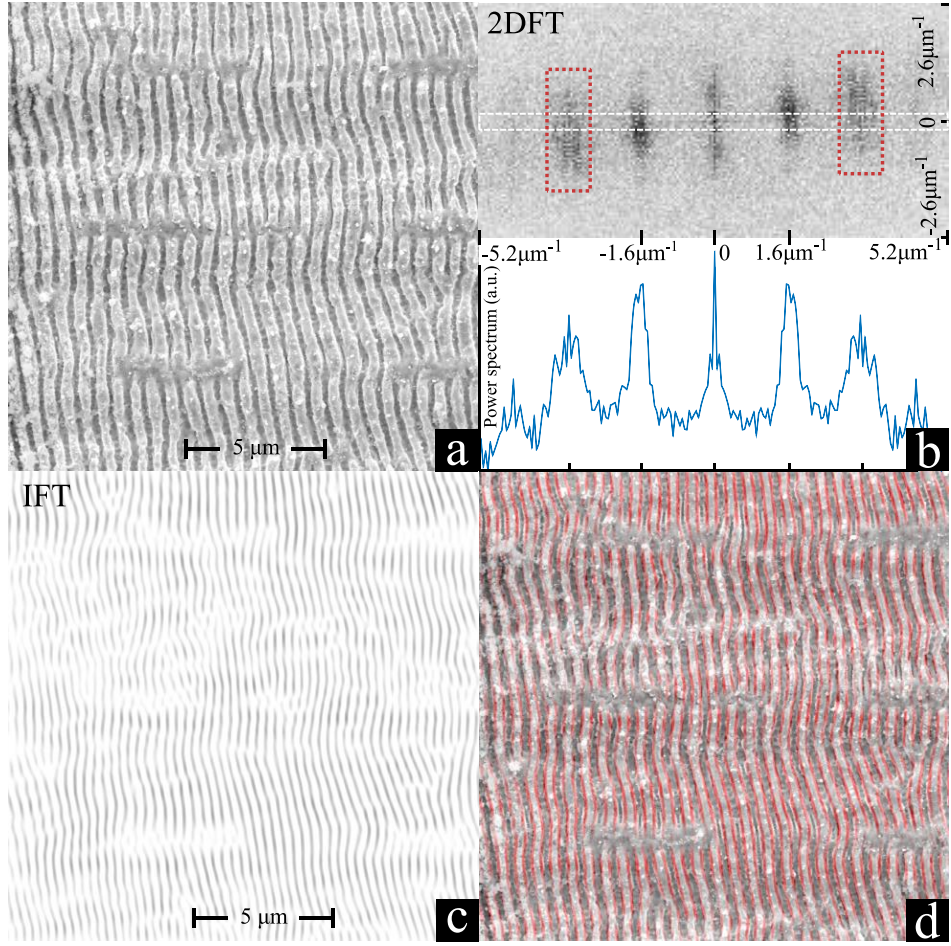

FIG. 3: SEM image and analysis of the inverse Fourier transform on the ~ ±3.2 μm$^{-1}$ frequencies to verify that the area with double the frequency of the ripples belongs to the edges of the ripples.

## IV. Heat conductivity of carriers

Heat conductivity of the carriers (denoted with 'c', where it is 'e' ('h'), for electrons (holes), respectively) is provided by the following expressions

$$k_c \simeq \frac{4k_B^2 T_c \sigma_c}{e}$$

$$\sigma_c \simeq eN_c \mu_c^0$$

$$N_c = 2\left[\frac{m^*_{c-cond} k_B T_c}{2\pi \hbar^2}\right]^{3/2} F_{1/2}(\eta_c)$$

where $F_{1/2}(\eta_c)$ are Fermi-Dirac integrals of order ½ ([32,59,60]). The approximating values for the first two equations are due to the fact that a Maxell-Boltzmann distribution for the carriers is assumed (non-degeneracy). To compare the electron (hole) conductivity for Silicon and GaAs, we take into account the carrier mobilities $\mu_c^0$ and their optical effective masses for the two materials: $\mu_e^0$ (GaAs) = 8500 cm$^2$/Vs, $\mu_h^0$ (GaAs) = 400 cm$^2$/Vs, $\mu_e^0$ (Si) = 1400 cm$^2$/Vs, $\mu_h^0$ (Si) = 450 cm$^2$/Vs, $m_{e-cond}^*$ (GaAs) = 0.067 $m_{e0}$, $m_{h-cond}^*$ (GaAs) = 0.34 $m_{e0}$, $m_{e-cond}^*$ (Si) = 0.33 $m_{e0}$, $m_{h-cond}^*$ (Si) = 0.81 $m_{e0}$ ([60,62]). These values yield the following ratios $\frac{k_e(\text{GaAs})}{k_e(\text{Si})} \simeq 0.55$ and $\frac{k_h(\text{GaAs})}{k_h(\text{Si})} \simeq 0.24$ which suggests that the carrier conductivity for GaAs is smaller (or at least of the order) than that of the Si. These results indicate that carrier diffusion does not have a substantial impact on the electron-hole creation as in Si [60].

## V. Computation of carrier collision time

An alternative methodology to compute the carrier collision time [60] was by considering that the electron-phonon, hole-phonon and electron-hole collisions contribute to the total collision frequency of the particles [60]. In those reports, the electron-phonon and hole-phonon collision frequencies are assumed to be identical and computed through the empirical expression $A_{eh}T_L$ ($A_{eh} = 1 \times 10^{11}$ s$^{-1}$K$^{-1}$); by contrast, the electron-hole collision time is estimated through the expression (that is valid for a non-degenerate carrier system) [60]

$$1/\tau_{e-h} = \frac{\sqrt{3}\varepsilon_0 \pi (k_B T_e)^{3/2}}{2e^2} \left[ \frac{1}{m_{e-DOS}^*} + \frac{1}{m_{h-DOS}^*} \right]^{1/2}$$

where $m_{e-DOS}^*$ = 0.067 $m_{e0}$ and $m_{h-DOS}^*$ = 0.47 $m_{e0}$ [67]. Similarly, Time Dependent Density Functional Theories could be used to provide an even more accurate description of the carrier dependent dielectric constant. Nevertheless, in the current work, a simplistic approach (similar to the one used for Silicon that appears to describe damage thresholds and carrier dynamics efficiently) was followed where $\tau_{col}$ was assumed to be constant.

## VI. Height of ripples *vs* number of pulses and temperature field distribution

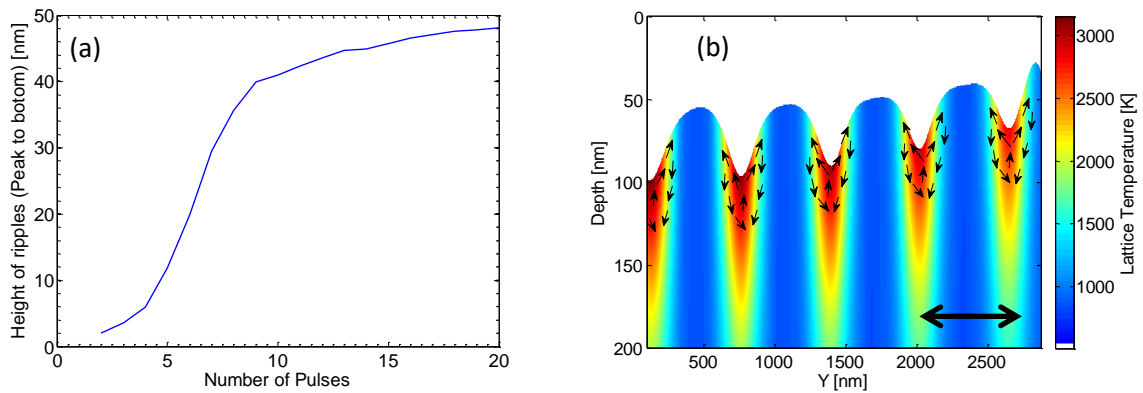

FIG. 4: (a) Height of ripples as a function of *NP*. (b) Spatial distribution of Lattice temperature at $t=$ 10 ns for *NP*=14 along the *Y*-direction ($E_p$=200 mJ/cm$^2$, $\tau_p$=100 fs). *Black Arrows* indicate the fluid movement. The double-arrow represents polarization of the laser beam.

## VII. Ablation conditions

In principle, femtosecond pulsed laser interaction with matter triggers a variety of timescale-dependent processes, influenced by the fluence and pulse duration; different combinations of those parameters are capable to induce phase transition or material removal. A solid material subjected to ultrashort pulsed laser heating at sufficiently high fluences undergoes a phase transition to a superheated liquid whose temperature reaches $0.90 T_{cr}$ ($T_{cr}$ being the thermodynamic critical temperature). A subsequent bubble nucleation leads to a rapid transition of the superheated liquid to a mixture of vapour and liquid droplets that are ejected from the bulk material (phase explosion). This has been proposed as a material removal mechanism. By contrast, the interpretation of a possible surface modification due to evaporation has been related to the presence of a Knudsen layer adjacent to the liquid-vapor interface and the process has been analysed in numerous works. The proposed scenario of modelling material removal is based on a combination of evaporation of material volumes that exceed upon irradiation lattice temperatures close to $0.90 T_{cr}$ and evaporation due to dynamics of Knudsen layer, which is a scenario that was elaborated in a previous work [20].

On the other hand, as values for $T_{cr}$ or $T_b$ (boiling point temperature) are not known for GaAs, based on experimental observations for mass removal (i.e. ablation) for Silicon [20] and Steel [15], $T_{rem}$ (the simulated lattice temperature for which the experimental conditions yielded onset of ablation) was computed to be about $\sim 3 \times T_{melt}$ while $T_b$ again is approximately taken to be around $\sim 2.5 \times T_{melt}$. Although this value is rather approximate, however, it is unlikely that a more detailed methodology of estimating the amount of ablated region (and the temperatures that lead to ablation) would change the picture substantially due to the small height of the ablated region (~less than 2-3 nanometers for each pulse).